\documentstyle[prl,aps,epsfig,floats]{revtex}
\def\be{\begin{eqnarray}}
\def\ee{\end{eqnarray}}
\def\ben{\begin{equation}}
\def\een{\end{equation}}

\def\etal{{\it et al.}}
\newcommand{\e}{{\mbox{e}}}

\def\vA{{\bbox  A}}

\def\vV{{\bbox  V}}
\def\vr{{\bbox  r}}

\def\vq{{\bbox  q}}
\def\vk{{\bbox  k}}
\def\vp{{\bbox  p}}

\def\vx{{\bbox  x}}

\def\vs{{\bbox \sigma}}

\def\vbp{{\overline \vp}}
\def\hatr{{\bbox {\hat r}}}

\def\He#1{{}^{#1}\mbox{He}}

\def\nlo#1{\mbox{N$^{#1}$LO}}
\def\Sunit{\mbox{$10^{-20}$ keV-b}}
\def\voO#1{{\bbox {\cal O}}_{\!#1}}
\def\voT#1{{\bbox {\cal T}}_{\!\!#1}}
\def\dR{{\hat d^R}}
\def\bi{\bibitem}
\def\prl{Phys. Rev. Lett.}\def\pr{Phys. Rev.}
\def\pl{Phys. Lett.}

\begin{document}

\twocolumn[%
\hsize\textwidth\columnwidth\hsize\csname@twocolumnfalse\endcsname

\renewcommand{\thefootnote}{\fnsymbol{footnote}}
\setcounter{footnote}{0}

\title{%
\bf The Solar $hep$ Process in Effective Field Theory}

\author{%
{\bf T.-S. Park}$^{(a)}$, {\bf  L.E. Marcucci}$^{(b,c)}$, {\bf R.
Schiavilla}$^{(d,e)}$, {\bf M. Viviani}$^{(c,b)}$, {\bf A.
Kievsky}$^{(c,b)}$, {\bf S. Rosati}$^{(b,c)}$
\\
{\bf K. Kubodera}$^{(a)}$, {\bf D.-P. Min}$^{(f)}$, and {\bf M. Rho}$^{(f,g,h)}$}
\address{%
(a) Department of Physics and Astronomy,
University of South Carolina, Columbia, SC 29208, USA\\
(b) Department of Physics, University of Pisa, I-56100 Pisa,
Italy\\
(c) INFN, Sezione di Pisa, I-56100 Pisa, Italy\\
(d) Department of Physics, Old Dominion University,
         Norfolk, Virginia 23529, USA \\
(e) Jefferson Lab, Newport News, Virginia 23606, USA \\
(f) School of Physics and Center for Theoretical Physics,
Seoul National University, Seoul 151-742, Korea\\
(g) Service de Physique Th\'{e}orique, CEA  Saclay, \it
91191 Gif-sur-Yvette Cedex, France\\
(h) Institute of Physics and Applied Physics, Yonsei University,
Seoul 120-749, Korea
}

\maketitle

\centerline{(July 24, 2001)}

\begin{abstract}
Using effective field theory,
we calculate the $S$-factor for the $hep$ process 
in a totally parameter-free formulation.
The transition operators are organized 
according to chiral counting, and their
matrix elements are evaluated 
using the realistic nuclear wave functions obtained 
in the correlated-hyperspherical-harmonics method. 
Terms of up to next-to-next-to-next-to-leading order in
heavy-baryon chiral perturbation theory are considered. 
Fixing the only parameter in the theory 
by fitting the tritium $\beta$-decay
rate, we predict the $hep$ $S$-factor with accuracy 
better than $\sim 20$ \%.
\end{abstract}

 \vskip1pc]

\renewcommand{\thefootnote}{\arabic{footnote}}
\setcounter{footnote}{0}

The $hep$ process \be p+\He3 \rightarrow \He4 + e^+ + \nu_e
\label{hep} \ee figures importantly in astrophysics and particle
physics; it has much bearing upon issues of great current interest
such as the solar neutrino problem, non-standard physics in the
neutrino sector, etc. 
The $hep$ reaction produces the highest-energy
solar neutrinos even though their flux is much smaller than that
of the ${}^8\mbox{B}$ neutrinos. At some level, therefore, there
can be a significant distortion of the higher end of the
${}^8\mbox{B}$ neutrino spectrum due to the $hep$ neutrinos. 
This
change can influence the interpretation of the results of a recent
Super-Kamiokande experiment that have generated many controversies
related to the solar neutrino problem and neutrino
oscillations~\cite{controversy,monderen}. To address these issues,
a reliable estimate of the $hep$ cross section is indispensable.
Its accurate evaluation, however, presents a challenge for nuclear
and hadron physics~\cite{challenge}. The degree of this difficulty
may be appreciated by noting that theoretical estimates of the
$hep$ astrophysical $S$-factor have varied by orders of magnitude
in the literature. The reason for this great variation is
multifold. Although the primary $hep$ amplitude is formally of the
Gamow-Teller (GT) type ($\Delta J=1$, same parity), the
single-particle GT matrix element for the $hep$ process is
strongly suppressed due to the symmetries of the initial and final
state wave functions. Furthermore, the main two-body corrections
to the ``leading" one-body GT term generically come with opposite
sign causing a large cancellation. Therefore, it is necessary to
calculate these corrections with great accuracy, which is a highly
non-trivial task. The nature of the specific challenge involved
here can be further elucidated in terms of the {\it chiral filter}
picture. If meson-exchange current (MEC) contributions to a
specific transition amplitude are dominated by the one-soft-pion
exchange diagram, then one can take advantage of the fact that the
soft-pion-exchange piece is uniquely dictated by chiral symmetry
and that there is a mechanism (called the chiral filter mechanism)
that suppresses higher chiral-order terms~\cite{KDR,BR2001}. We
refer to a transition amplitude to which the chiral filter
mechanism is applicable (not applicable) as a chiral-protected
(chiral-unprotected) case. It is known that the space component of
the vector current and axial-charge are chiral-protected, whereas
the GT transition is chiral-unprotected. 
This means that, in
contrast to isovector M1 and axial charge transitions, both of
which are chiral-protected and hence easily
calculated~\cite{M1,axialch}, 
the primary $hep$ amplitude, i.e.
the GT transition, needs to be calculated without invoking the
chiral-filter mechanism and hence is more subtle.

In a highly successful method, which we call here
the standard nuclear physics approach (SNPA)~\cite{snpa},
one uses the potential picture supplemented
with MEC contributions,
the structures of which are obtained from one-boson-exchange.
Many applications of SNPA to electroweak
transitions in few-body systems are well documented.
A detailed SNPA calculation for the $hep$ cross section
was recently carried out by Marcucci {\etal}~\cite{MSVKRB};
this study has re-confirmed the substantial cancellation
between the one-body and two-body terms for the
GT transition~\cite{CRSW91,SWPC92}.
It is therefore of primary importance to
have a good theoretical control of short-distance physics.
A ``first-principle" approach based on
effective field theory (EFT) will hopefully provide an insight
into this issue.
A possible approach that is formally consistent
with systematic power counting is the pionless EFT
(for a recent review, see \cite{seattle}).
This approach, however, cannot be
readily extended to systems with $A\ge3$.
Apart from the basic problem of organizing chiral expansion
for complex nuclei from ``first-principles",
a plethora of parameters involved would present
a major obstacle.
This difficulty is expected to be
particularly pronounced for the $hep$ reaction.

In this paper we calculate the $hep$ astrophysical $S$-factor,
adopting a variant of EFT, that has been proven to be extremely
successful in estimating the $S$-factor for solar $pp$
fusion~\cite{PMetal2001}. The same method has also scored great
success in predicting polarization observables in thermal $np$
capture~\cite{PKMRnp}. Our starting point is the observation that,
to high accuracy, the leading-order single-particle matrix
elements in SNPA and EFT are identical, and that they can be
reliably estimated with the use of realistic SNPA wave functions
for the initial and final nuclear states. Next, we note that in
EFT the operators representing two-body corrections to the
leading-order one-body term can be controlled by systematic chiral
expansion in heavy-baryon chiral perturbation theory~\cite{PKMR}.
Then, since the ratio of a two-body matrix element to the
leading-order one-body matrix element can be evaluated with
sufficient accuracy with the use of the realistic SNPA wave
functions, we are in a position to obtain a reliable estimate of
the total (one-body + two-body) contribution. This method, which
exploits the powers of {\it both} SNPA and EFT, may be referred to
as a ``{\it more effective} EFT (MEEFT)". Here we shall work out
MEEFT up to next-to-next-to-next-to-leading order (\nlo3),
adopting a cut-off regularization. Our method, which to this order
preserves chiral symmetry and renormalization group invariance,
enables us to predict the $hep$ $S$-factor in a totally
parameter-free manner.

In the present scheme it is sufficient to focus on
``irreducible graphs" in Weinberg's
classification~\cite{weinberg}.
Graphs are classified by the
chiral power index $\nu$ given by
$\nu = 2 (A-C) + 2 L +\sum_i \nu_i \label{nu}$,
where $A$ is the number of nucleons involved
in the process ($A$=4 in our case),
$C$ the number of disconnected parts,
and $L$ the number of loops.
The chiral index, $\nu_i$, of the $i$-th vertex is given by
$\nu_i= d_i + e_i + n_i/2-2 \label{nui}$,
where $d_i$, $e_i$ and $n_i$ are respectively
the numbers of derivatives, external fields
and nucleon lines belonging to the vertex.
The Feynman diagrams with a chiral index $\nu$
are suppressed by $(Q/\Lambda_\chi)^\nu$
compared with the leading-order one-body GT operator,
with $Q$ standing for the typical three-momentum scale
and/or the pion mass, and
$\Lambda_\chi \sim m_N \sim 4\pi f_\pi$
for the chiral scale.
The physical amplitude is
then expanded with respect to $\nu$.

In this paper we shall limit ourselves to \nlo3,
although it is possible to go to \nlo4
without introducing new parameters.
We write the current as
 \be J^\mu(\vq)= V^\mu(\vq) + A^\mu(\vq)
 = \int d\vx\, \e^{- i\vq\cdot \vx} J^\mu(\vx),
\ee
where $\vq$ is the momentum carried by the lepton pair.
The calculation is considerably simplified
by the facts that we are dealing
with small momentum transfers
and that, as a consequence,
both $q\equiv|\vq|$ and the time component
of the nucleon momentum are
${\cal O}(Q^2/\Lambda_\chi)$
rather than ${\cal O}(Q)$
as naive counting would suggest.
For the vector and axial
one-body current and charge operators, 
and also for the vector-current and axial-charge 
(chiral-protected) two-body operators,
we can simply carry over the expressions 
given in Ref.~\cite{MSVKRB},
hereafter referred to as MSVKRB.
The vector charge two-body operator does not appear
up to the order considered, while
the EFT axial-current two-body operator
is given, in momentum space, in Ref.~\cite{PMetal2001}.
These two-body currents are valid
only up to a cutoff $\Lambda$.
This implies that, when we go to coordinate space,
the currents must be regulated.
This is a key point in our approach.
We introduce a Gaussian cutoff regulator
in performing Fourier transformation.
The resulting $r$-space expressions of the currents
in the center-of-mass (c.m.) frame are
\be
\vV_{12}(\vr) &=& - 
\frac{g_A^2 m_\pi^2}{12 f_\pi^2}
\tau_\times^- \,
 \vr \,
 \left[
 \vs_1\cdot\vs_2 \, y_{0\Lambda}^\pi(r)
 + S_{12} \, y_{2\Lambda}^\pi(r) \right]
\nonumber \\
 &-& i \frac{g_A^2}{8 f_\pi^2}
 \vq\times \left[
 \voO\times y_{0\Lambda}^\pi(r)
 +\left( \voT\times - \frac23 \voO\times \right)
  y_{1\Lambda}^\pi(r)
 \right],
\nonumber \\
A^{0}_{12}(\vr) &=& -
\frac{g_A}{4 f_\pi^2}
\tau_\times^- 
\left[
\frac{\vs_+ \cdot \hatr}{r}
+
\frac{i}{2} \vq\cdot \hatr\, \vs_-\cdot \hatr\,
\right]
y_{1\Lambda}^\pi(r) ,
\nonumber \\
\vA_{12}(\vr) &=&
- \frac{g_A m_\pi^2}{2 m_N f_\pi^2}
\Bigg[  
\nonumber \\
&&\left[
\frac{\hat c_3}{3} (\voO+ + \voO-)
+\frac23 \left(\hat c_4 + \frac14\right)
   \voO\times \right] y_{0\Lambda}^\pi(r)
\nonumber \\
&&
  + \left[
     \hat c_3 (\voT+ + \voT-)
  - \left(\hat c_4 + \frac14\right) \voT\times
     \right] y_{2\Lambda}^\pi(r)
   \Bigg]
\nonumber \\
&+& \frac{g_A}{2 m_N f_\pi^2 }
\Big[
\frac{1}{2} \tau_\times^-
   (\vbp_1 \,\vs_2\cdot\hatr +
   \vbp_2\,\vs_1\cdot\hatr)
\frac{y_{1\Lambda}^\pi(r)}{r}
\nonumber \\
&&
 + \delta_\Lambda(r)
 \, \hat d^R \voO\times
\Big],
\label{vAnuFT}\ee
where $\vr=\vr_1 - \vr_2$, $S_{12}$ is
the tensor operator, and
\be
\left[\delta_\Lambda(r),
\ y_{0\Lambda}^\pi(r)\right] \equiv
 \int\!\!\frac{d\vk}{(2\pi)^3}\,
 \e^{-k^2/\Lambda^2}
\e^{i \vk\cdot \vr}
\left[1,\ \frac{1}{\vk^2 + m_\pi^2}\right], \nonumber \ee
$y_{1\Lambda}^\pi(r) \equiv - r
[y_{0\Lambda}^\pi(r)]^\prime$
and
$y_{2\Lambda}^\pi(r) \equiv
(r/m_\pi^2) [ [y_{0\Lambda}^\pi(r)]^\prime/r]^\prime$
(the $^\prime$ symbols denote derivatives),
$\voO{\odot}^{k} \equiv \tau_\odot^- \sigma_\odot^k$,
$ \voO{\odot} \equiv \tau_\odot^- \vs_\odot$,
$ \voT{\odot} \equiv
 \hatr\, \hatr\cdot \voO{\odot} - \frac13 \voO{\odot}$,
$\odot=\pm,\times$,
$\tau_\odot^-\equiv(\tau_1\odot\tau_2)^-\equiv
(\tau_1\odot\tau_2)^x -i (\tau_1\odot\tau_2)^y$ and
$\vs_\odot\equiv(\vs_1\odot\vs_2)$.
Finally, the parameter $\dR$ is given by
\be \dR\equiv \hat d_1 +2 \hat d_2 +
\frac13 \hat c_3
 + \frac{2}{3} \hat c_4 + \frac16\,.
\ee
The dimensionless parameters, $\hat c$'s and $\hat d$'s,
are defined in Ref.~\cite{pkmr-apj}.
The numerical values of
$\hat c_3$ and $\hat c_4$ have been determined
previously~\cite{csTREE}: $\hat c_3 = -3.66 \pm 0.08$
and $\hat c_4 = 2.11 \pm 0.08$.
The derivative operators, $\vbp_i \equiv \frac12 (\vp_i' + \vp_i)$
($i=1,\,2$) in Eq.(\ref{vAnuFT}), should be understood to act only on the
wave functions.

The explicit degrees of freedom in our scheme
are the nucleon and the pion,
with all other degrees of freedom
($\rho$- and $\omega$-mesons, $\Delta (1232)$, etc.)
integrated out.
Therefore, a reasonable range of the cutoff $\Lambda$
would be somewhere around 500--800 MeV.
We shall consider here three exemplary values,
$\Lambda=500,\ 600,\ 800$ MeV.

A crucial point is that $\vA_{12}$ in Eq.(\ref{vAnuFT})
contains only one unknown parameter, $\hat{d}^R$, that needs to be
fixed using an empirical input. The tritium $\beta$-decay rate,
$\Gamma_\beta$, can be used for this purpose, since it is
dominated by the {\it unsuppressed} GT term \cite{TBDexp}. Thus,
for each value of $\Lambda$, we adjust $\dR$ to reproduce the
experimental value of $\Gamma_\beta$ and, using the value of $\dR$
so determined, we evaluate the $hep$ amplitude. To carry out this
program, we must calculate both the tritium $\beta$-decay
amplitude and the $hep$ amplitude for the above-described
currents, using realistic wave functions. We adopt here the method
employed by MSVKRB. In particular, we use the
correlated-hyperspherical-harmonics wave
functions~\cite{VKR95,VRK98} obtained with
the Argonne $v_{18}$ (AV18) two-nucleon~\cite{av18} and Urbana-IX
three-nucleon~\cite{uix} interactions.

In the notation of MSVKRB, the GT-amplitudes are given in terms of
the reduced-matrix elements (RMEs) $\overline{L}_1(q;A)$ and
$\overline{E}_1(q;A)$. Since these RMEs are related to each other
as $\overline{E}_1(q;A) \simeq \sqrt{2}\, \overline{L}_1(q;A)$,
with the exact equality holding at $q$=0, we consider here only
one of them, $\overline{L}_1(q;A)$. For the three exemplary values
of $\Lambda$, Table~\ref{TabL1A} gives the corresponding values of
$\dR$, as determined from $\Gamma_\beta$~\cite{PMetal2001}, and
$\overline{L}_1(q;A)$ at $q$=19.2 MeV and zero c.m.
energy.  We see from the table that
the variation of the two-body GT amplitude (row labelled
``2B-total'') is only $\sim$10 \% for the range of $\Lambda$ under
study. The $\Lambda$-dependence in the total GT amplitude is made
more pronounced by a strong cancellation between the one-body
and two-body terms, but this amplified $\Lambda$-dependence still
lies within acceptable levels.

\begin{table}[htb]
\centering
\begin{tabular}{c|rrr}
$\Lambda$ (MeV) & 500 & 600 & 800 \\ \hline
$\dR$      & $1.00 \pm 0.07$ & $1.78\pm 0.08$ & $3.90\pm 0.10$
\\ \hline
$\overline{L}_1(q;A)$ & $-0.032$ & $-0.029$ & $-0.022$ \\ \hline
1B                    & $-0.081$ & $-0.081$ & $-0.081$ \\ \hline
2B (without $\dR$)    & $0.093$  & $0.122$  & $0.166$  \\
2B ($\propto \dR$)    & $-0.044$ & $-0.070$ & $-0.107$ \\ \hline
2B-total              & $0.049$  & $0.052$  & $0.059$  \\
\end{tabular}
\caption{\label{TabL1A}\protect Values of $\dR$ and
$\overline{L}_1(q;A)$ (in fm$^{3/2}$)
calculated as functions of the cutoff $\Lambda$.
The individual contributions from the one-body (1B) and two-body (2B)
operators are also listed.
}
\end{table}

\begin{table}[hbt]
\centering
\begin{tabular}{c|ccc|c}
$\Lambda$ (MeV) & 500 & 600 & 800 & MSVKRB
\\ \hline
${}^1S_0$ & 0.02  &  0.02  &  0.02 & 0.02 \\
${}^3S_1$ & 7.00  &  6.37  &  4.30 & 6.38 \\
${}^3P_0$ & 0.67  &  0.66  &  0.66 & 0.82 \\
${}^1P_1$ & 0.85  &  0.88  &  0.91 & 1.00 \\
${}^3P_1$ & 0.34  &  0.34  &  0.34 & 0.30 \\
${}^3P_2$ & 1.06  &  1.06  &  1.06 & 0.97 \\ \hline
Total     & 9.95  &  9.37  &  7.32 & 9.64 \\ 
\end{tabular}
\caption{\label{TabS}\protect Contributions
to the $S$-factor (in \Sunit)
from individual initial channels calculated as functions of $\Lambda$.
The last column gives the results obtained in MSVKRB.}
\end{table}

Table~\ref{TabS} shows the contribution to the $S$-factor,
at zero c.m. energy, from each initial channel.
For comparison we have also listed the MSVKRB results
for the AV18/UIX interaction.
It is noteworthy that for all the channels
other than ${}^3S_1$,
the $\Lambda$-dependence is very small ($\lesssim 2$ \%).
While the GT terms are dominant,
the contribution of the axial-charge term
in the ${}^3S_1$ channel is sizable
even though it is kinematically suppressed
by the factor $q$.
It is therefore reassuring
that the chiral-filter mechanism allows
a reliable evaluation of this amplitude.

Summarizing the results given in Table~\ref{TabS}, we arrive at a
quantitative prediction for the $hep$ $S$-factor: 
\be S=(8.6 \pm 1.3 )\times \Sunit\,,\label{prediction} \ee 
where the ``error"
spans the range of the $\Lambda$-dependence for $\Lambda$=500--800
MeV. This result should be compared to that obtained by
MSVKRB~\cite{MSVKRB}, $S=9.64 \times \Sunit$.  Note that the
earlier studies~\cite{CRSW91,SWPC92} were based on less accurate
variational wave functions than used here and in MSVKRB, and did
not include P-wave capture contributions, which account for
$\simeq 40$ \% of the total $S$-factor. To decrease the
uncertainty in Eq.(\ref{prediction}), we need to reduce the
$\Lambda$-dependence in the two-body GT term. According to a
general {\it tenet} of EFT, the $\Lambda$-dependence should
diminish as we include higher order terms. A preliminary
study~\cite{PKMRhep} indicates that it is indeed possible to
reduce the $\Lambda$-dependence significantly by including \nlo4
corrections.

In order to better understand how the present scheme works,
it is helpful to compare the $hep$ reaction with the
radiative $np$-capture.
The polarization observables in
$\vec{n}+\vec{p}\rightarrow d+\gamma$ are known
to be sensitive to the isoscalar M1 matrix element,
$M1S$,
and this amplitude has been extensively studied
in EFT \cite{PKMRnp,crs99}.
The similar features of the $hep$ GT amplitude and the $M1S$ matrix element
are: (i) the leading one-body contribution
is suppressed by the symmetries of the wave functions;
(ii) there is no soft-pion exchange contribution;
(iii) nonetheless,
short-range physics can be reliably subsumed
into a single contact term.
In the $\vec{n}\vec{p}$ case
the strength of this term can be determined
from the deuteron magnetic moment
(for a given value of the cutoff $\Lambda$).
The calculation in Ref.~\cite{PKMRnp} demonstrates
that the $\Lambda$-dependence
in the contact term and that of the remaining terms
compensate each other so that
the total $M1S$ is stable against changes
in $\Lambda$.
This suggests that, if we go to higher orders,
the coefficient of the contact term
in question will be modified,
with part of its strength shifted to higher order terms;
however, the total physical amplitude will
remain essentially unchanged.
These features are quite similar to what we have found here
for the $hep$ GT amplitude.

Evaluating the matrix element of the leading-order
one-body operator in EFT with the use of
realistic nuclear wave functions is analogous
to fixing parameters in an EFT Lagrangian (at a given order)
using empirical inputs.
The realistic wave functions in SNPA can be regarded as a
theoretical input that fits certain sets of
observables~\cite{PKMR98}. In the present MEEFT scheme, we take
the view that the same realistic wave functions also provide a
framework for reliably calculating corrections to the
leading-order one-body matrix element. While from a formal point
of view the approach adopted here is, in certain cases, not in
strict accordance with the systematic power-counting scheme of EFT
proper, nevertheless the severity of this potential short-coming
may depend on individual cases (see discussion
in~\cite{beaneetal2}). For the $hep$ amplitude under consideration
here, the moderate $\Lambda$-dependence exhibited by the numerical
results suggests that the lack of rigorous power-counting cannot
be too significative. Indeed, this type of ``resilience" may also
explain why the SNPA calculation in Ref.~\cite{MSVKRB} gives a
result very similar to the present one. It is true that the MSVKRB
two-body terms are not entirely in conformity with the chiral
counting scheme we are using here; some terms corresponding to
chiral orders higher than \nlo3 are included, while some others
which are \nlo3 in EFT are missing. This formal problem, however,
seems to be largely overcome by the fact that also in MSVKRB a
parameter (the axial $\pi N\Delta$ coupling strength) is adjusted
to reproduce $\Gamma_\beta$.

The latest analysis of the Super-Kamiokande data
\cite{SK2001} gives an upper limit
of the solar $hep$ neutrino flux,
$\Phi(hep)^{\rm SK} < 40
\times 10^3$ cm$^{-2}$s$^{-1}$.
The standard solar model
\cite{BP2000} using the $hep$ $S$-factor
of Marcucci {\etal}\cite{MSVKRB} predicts
$\Phi(hep)^{\rm SSM} = 9.4
\times 10^3$ cm$^{-2}$s$^{-1}$.
The use of the central value of our estimate,
Eq.(\ref{prediction}), of the $hep$ $S$-factor
would slightly lower $\Phi(hep)^{\rm SSM}$
but with the upper limit compatible with
$\Phi(hep)^{\rm SSM}$ in Ref.~\cite{BP2000}.
A concrete 
estimate of the
theoretical uncertainty in Eq.(\ref{prediction})
is expected to be useful for further discussion of the
solar $hep$ problem.

The work of TSP and KK is supported in part
by the U.S. National Science Foundation,
Grant Nos. PHY-9900756 and INT-9730847, while
that of RS is supported by the U.S. Department 
of Energy contract DE-AC05-84ER40150
under which the Southeastern Universities 
Research Association (SURA)
operates the Thomas Jefferson National Accelerator Facility.
The work of DPM is supported in part 
by KOSEF Grant 1999-2-111-005-5 and 
KSF Grant 2000-015-DP0072.
MR acknowledges the hospitality of the Physics
Departments of Seoul National University and Yonsei University,
where his work was partially supported by Brain Korea 21 in 2001.

\thebibliography{99}

\bi{controversy} 
J.N. Bahcall, Phys. Repts. {\bf 333}, 47 (2000) and references
given therein.

\bi{monderen} R. Escribano, J.-M. Frere, A. Gevaert, and D.
Monderen, Phys. Lett. {\bf B444}, 397 (1998).

\bi{challenge} J.N. Bahcall, hep-ex/0002018; J.N. Bahcall and P.I.
Krastev, Phys. Lett. {\bf B436}, 243 (1998).

\bi{KDR} K. Kubodera, J. Delorme, and M. Rho, \prl \ {\bf 40}, 755
(1978);  M. Rho, \prl\ {\bf 66}, 1275 (1991).

\bi{BR2001} G.E. Brown and M. Rho,  hep-ph/0103102, Physics
Reports, in press.

\bibitem{M1} T.-S. Park, D.-P. Min, and M. Rho,
Phys. Rev. Lett. {\bf 74}, 4153 (1995); Nucl. Phys. {\bf A596},
515 (1996).

\bibitem{axialch} T.-S. Park, D.-P. Min, and M. Rho,
Phys. Repts. {\bf 233}, 341 (1993);
T.-S. Park, I.S. Towner, and K. Kubodera,
Nucl. Phys. {\bf A579}, 381 (1994).

\bi{snpa} For a recent review, see J. Carlson and R. Schiavilla,
Rev. Mod. Phys. {\bf 70}, 743 (1998).

\bibitem{MSVKRB} L.E. Marcucci {\it et al.},
Phys. Rev. {\bf C63}, 015801 (2001).

\bibitem{CRSW91}
J. Carlson, D.O. Riska, R. Schiavilla, and R.B. Wiringa, Phys. Rev.
{\bf C44}, 619 (1991).

\bibitem{SWPC92} R. Schiavilla, R.B. Wiringa, V.R. Pandharipande, and
J. Carlson, Phys. Rev. {\bf C45}, 2628 (1992).

\bi{seattle} S.R. Beane {\it et al.}, nucl-th/0008064.

\bibitem{PMetal2001} T.-S. Park {\it et al.},
nucl-th/0106025, submitted to Phys. Rev. Lett.

\bi{PKMRnp}
T.-S. Park, K. Kubodera, D.-P. Min, and M. Rho, \pl \
{\bf B472}, 232 (2000).

\bibitem{PKMR}
T.-S. Park, K. Kubodera, D.-P. Min, and M. Rho,
Nucl. Phys. {\bf A684}, 101 (2001); see also nucl-th/9904053.

\bibitem{weinberg} S. Weinberg, Phys. Lett. {\bf B251}, 288 (1990);
Nucl. Phys. {\bf B363}, 3 (1991).

\bibitem{pkmr-apj}
T.-S. Park, K. Kubodera, D.-P. Min, and M. Rho,
Astrophys. J. {\bf 507}, 443 (1998).

\bibitem{csTREE} V. Bernard, N. Kaiser and Ulf.-G. Mei{\ss}ner,
Nucl. Phys. {\bf B457}, 147 (1995).

\bibitem{TBDexp} R. Schiavilla {\etal},
Phys. Rev. {\bf C58}, 1263 (1998).

\bibitem{VKR95} M. Viviani, A. Kievsky, and S. Rosati,
Few-Body Syst. {\bf 18}, 25 (1995).

\bibitem{VRK98} M. Viviani, S. Rosati, and A. Kievsky,
Phys. Rev. Lett. {\bf 81}, 1580 (1998).

\bibitem{av18}
R.B. Wiringa, V.G.J. Stoks, and R. Schiavilla,
Phys. Rev. {\bf C51}, 38 (1995).

\bibitem{uix}
B.S. Pudliner, V.R. Pandharipande, J. Carlson, and R.B. Wiringa,
Phys. Rev. Lett. {\bf 74}, 4396 (1995).

\bi{PKMRhep} T.-S. Park, K. Kubodera, D.-P. Min, and M. Rho,
unpublished.

\bi{crs99}
J.-W. Chen, G. Rupak, and M.J. Savage, \pl \
{\bf B464}, 1 (1999).

\bi{PKMR98} T.-S. Park, K. Kubodera, D.-P. Min, and M. Rho, \pr \
{\bf C58}, 637 (1998).

\bi{beaneetal2} S.R. Beane, P.F. Bedaque, M.J. Savage, and U. van
Kolck, nucl-th/0104030.

\bi{SK2001}
S. Fukuda {\it et al.},
Phys. Rev. Lett. {\bf 86}, 5651 (2001).

\bi{BP2000}
J.N. Bahcall {\it et al.}, astro-ph/0010346.
\end{document}